\documentstyle[prb,eqsecnum,aps]{revtex}

\begin{document}
\draft
\preprint{}
\widetext
\title{Magnetic Incommensurability in Doped Mott Insulator}
\author{ Z.Y. Weng, D.N. Sheng, and C.S. Ting }
\address{Texas Center for Superconductivity and Department of Physics\\
University of Houston, Houston, TX 77204-5506}
\maketitle
\begin{abstract}
In this paper we explore the incommensurate spatial modulation of spin-spin
correlations as the intrinsic property of the doped Mott insulator, described by
the $t-J$ model. We show that such an incommensurability is a direct 
manifestation of the phase string effect introduced by doped holes
in both one- and two-dimensional cases. The magnetic incommensurate peaks of 
dynamic spin susceptibility in momentum space are in agreement with the 
neutron-scattering measurement of cuprate superconductors in both position and
doping dependence. In particular, this incommensurate structure can naturally 
reconcile the neutron-scattering and NMR experiments of cuprates.

\end{abstract}


\widetext

\section{INTRODUCTION}

Antiferromagnetic (AF) spin correlations in cuprates persist into metallic phase
as evidenced, for example, in neutron-scattering measurements. Compared 
to insulating phase, however, a qualitative change in the nature of AF 
correlations has also been observed.\cite{inc1,inc2,inc3,inc4,inc5} 
The original magnetic peak in dynamic spin susceptibility centered at momentum 
$(\pi, \pi)$ splits into four incommensurate peaks at  $(\pi\pm\pi\bar{\delta},\pi)$ and
$(\pi, \pi\pm\pi\bar{\delta})$ in metallic or superconducting regime, as has 
been found in the $LSCO$ compounds\cite{inc1,inc2,inc3,inc4} (a recent
measurement\cite{inc5} indicates the similar structure for $YBCO$ 
compounds as well). The incommensurability are 
insensitive to both energy and temperature, corresponding to a new 
spatial modulation of spin-spin correlations in real space. 

The incommensurability $\bar{\delta}$ has been identified to be proportional to
doping concentration $\delta$ in approximately a {\it universal} fashion:
$\bar{\delta}\simeq 2\delta$.\cite{inc2,inc4} This observation poses a critical 
challenge to theories of incommensurability based on the Fermi-surface 
topology:\cite{fermi1,fermi2,fermi3}
since the Fermi-surface encloses a $1-\delta$ amount of electrons according to
the Luttinger theorem, it is not a trivial issue why the doping 
concentration $\delta$ of {\it holes} should dictate the
incommensurability. A very similar issue has already been well known in 
the transport properties of cuprates where charge carriers are always 
characterized by hole-nature with the total number determined by $\delta$ 
despite a large Fermi surface. This phenomenon has been interpreted due to 
the existence of a doped-Mott-insulator\cite{anderson} because of the strong 
Coulomb interaction among electrons. With striking similarity, magnetic 
incommensurability may be another intrinsic consequence of the doped-Mott-insulator.

A further challenge to any theory of magnetic incommensurability comes from
NMR spin relaxation rates, $1/T_1$.
As pointed out by Walstedt, Shastry, and Cheong,\cite{wsc} if the non-Korringa behavior of
$1/T_1$ for planar $^{63}Cu$ nuclear spins --- a hallmark of 
normal-state anomalies of cuprates --- originates from the AF correlations, this 
effect may generally `leak' to the planar $^{17}O$ nuclear spins if the AF 
correlations become incommensurate.  Thus the non-Korrigna behavior would have to
show up in the temperature dependence of $1/^{17}T_1$. 
This was indeed predicted by theories\cite{fermi1,fermi2,fermi3} where the incommensurability is 
interpreted in terms of the Fermi-surface topology.  But
experimentally the `leakage' of non-Korringa behavior to $^{17}O$ spins has 
never been observed,\cite{wsc} and the temperature behavior of $1/^{17}T_1$ is drastically different from that of $1/^{63}T_1$, implying a 
near-perfect {\it commensurate} AF correlations within a one-component-theory 
framework. So how to reconcile the NMR and neutron measurements remains to be a
serious puzzle in constructing a consistent microscopic theory. 

In this paper, we present a theoretical understanding of spatially  
modulated spin-spin correlations as a result of the doped Mott insulator 
described by the $t-J$ model. Such a Mott insulator at half-filling is an
antiferromagnet whose magnetic properties have been well understood.\cite{ins,ins1}
A hole doped into such an antiferromagnet will induce a phase string `defect' in the spin background which is nonrepairable at low-energy.\cite{string} It represents a 
singular doping effect and profoundly modifies both spin and charge 
correlations in the presence of finite density of doped holes.\cite{string1} 
At low-doping where the `rigidity' of the half-filled AF state still persists, 
described by a bosonic resonating valence bond (RVB) pairing,\cite{rvb,aa} a  
mean-field theory\cite{string2} incorporating the phase string 
effect is able to combine the AF insulating phase, 
superconducting phase, and an anomalous normal state within a unified 
phase diagram naturally. In the present paper we demonstrate that
it is in this regime a spatial modulation of spin-spin correlations indeed 
occurs as a direct manifestation of the nonrepairable phase string effect.  
Such an incommensurability is consistent with the experimental 
measurements in cuprate superconductors. In particular, such a mechanism 
for spatial modulation can uniquely reconcile the above-mentioned 
neutron and NMR experiments in cuprates. 

The present work will be based on the so-called phase string 
formalism\cite{string1} of the 
$t-J$ model in which the phase string effect is explicitly incorporated into
the Hamiltonian. In this formalism, spin raising operator $S^+_i$ reads
\begin{equation}\label{s+}
S^{+}_i=b^{\dagger}_{i\uparrow}b_{i\downarrow}(-1)^ie^{i\Phi_i^h}.
\end{equation}
Here $b_{i\sigma}$ denotes a bosonic spinon annihilation operator.
Compared to the 
conventional Schwinger-boson formalism,\cite{aa} a distinctive feature is 
the phase $\Phi_i^h$ which is defined by
\begin{equation}\label{phih}
\Phi_i^h\equiv \sum_{l\neq i}\mbox{Im ln $(z_i-z_l)$}  n_l^h. 
\end{equation}
It describes a phase vortex structure centered on holons:
$n_l^h$ is the holon number operator at site $l$ and
$z_i=x_i+iy_i$ represents the complex coordinate of a lattice site $i$.
If one changes $z_i$ continuously around a holon site $l$ once,  
$\Phi_i^h$ will gain an additional $2\pi$ phase. Such a vorticity around
a holon reflects the topology of the phase string effect
as schematically illustrated in Fig. 1. In this way, spins become nonlocally 
dependent on the positions of holes and the doping effect explicitly
enters spin-spin correlations through $\Phi^h_i$, in addition to a
modification of spinon spectrum via the Hamiltonian. 

In Sec. II, we show that an incommensurate spatial oscillation because of 
$\Phi^h_i$ will modulate the original commensurate oscillation of $(-1)^i$
in (\ref{s+}) at finite doping in both one-dimensional (1D) and two-dimensional 
(2D) cases. The incommensurability in 1D recovers the well-known Luttinger 
liquid behavior, while in 2D it explains the experimental results in cuprates 
including the incommensurate peak positions in momentum space and the doping
dependence of the incommensurability. We then further show that the effect of
such a spatial modulation of spin-spin correlations cannot be picked up by NMR
spin relaxation rates precisely due to its singular nature associated with doped 
holes, and thus reconcile the neutron and NMR experiments in cuprates. 
Finally, a conclusive discussion is presented in Sec. III.

\section{INCOMMENSURABILITY DUE TO PHASE-STRING EFFECT}

According to (\ref{s+}), the transverse spin-spin correlation function can be 
written down as follows 
\begin{eqnarray}\label{s+-}
\left\langle S^{+}_i(t) S^{-}_j(0)\right\rangle&=&(-1)^{i-j}
\left\langle b^{\dagger}_{i\uparrow}(t)b_{i\downarrow}(t)
\Pi^h_{ij}(t) b^{\dagger}_{j\downarrow}(0)b_{j\uparrow}(0)\right\rangle,
\end{eqnarray}
where
\begin{equation}
{\Pi}^h_{ij}(t)\equiv e^{i\Phi^h_i(t)}e^{-i\Phi^h_j(0)}.
\end{equation}
(The longitudinal correlation function will be discussed in the Appendix.)
At half-filling, $\Phi^h_i=0$ and the prefactor $(-1)^{i-j}$ in (\ref{s+-})  
determines a commensurate AF oscillation of the spin-spin correlation function: 
$(-1)^{i-j}\equiv e^{i{\bf Q}_0\cdot ({\bf r}_i-{\bf r}_j)}$ with the AF
wavevector 
\begin{equation}
{\bf Q}_0=(\pm \pi/a, \pm\pi/a),
\end{equation}
where $a$ is the lattice constant.
[Note that the average in (\ref{s+-}) will also produce a (small) term which oscillates as $(-1)^{i-j}$
in real space and compensates the prefactor in (\ref{s+-}), leading to a
ferromagnetic component of spin-spin correlations at half-filling.\cite{aa}]

At finite doping, the topological phase $\Phi^h_i$ defined in (\ref{phih}) will 
qualitatively modify the spatial modulation of the AF spin-spin correlations. 
For the purpose of illustration, in the following we discuss the 1D case 
first as an example, where the phase string effect has been already 
shown\cite{string1} to play a 
crucial role responsible for Luttinger liquid behavior. Then we discuss the 2D
case and make comparison with experiment.

\subsection{Incommensurability in 1D}

In 1D case, $\Phi^h_i$ simply reduces to\cite{string1}
\begin{equation}\label{ephh}
\Phi^h_i=\pm\pi\sum_{l>i} n_l^h .
\end{equation}
Let us consider the equal-time case first, where
\begin{equation}\label{phase1d}
\Pi^h_{ij}=e^{i(\Phi^h_i-\Phi^h_j)}=e^{\pm i\pi\sum_{i<l<j} n_l^h }.
\end{equation}
Here we assume $i<j$  without loss of generality.

Expression (\ref{phase1d}) clearly shows that every holon sitting between $i$ and 
$j$ will contribute a $\pi$ phase shift to $\Pi^h_{ij}$. Due to the spin-charge separation in 1D, one may take separate averages over spinon and holon in (\ref{s+-}). Define $N^h_{ij}=\sum_{i<l<j}
n^h_l$. The average of $\Pi^h_{ij}$ over holon degrees of freedom can be evaluated as follows
\begin{eqnarray}\label{ff}
\langle \Pi^h_{ij}\rangle &=&  e^{\pm i\pi \langle N_{ij}^h\rangle} \left\langle e^{\pm 
i\pi(N_{ij}^h-\langle N_{ij}^h\rangle)}\right\rangle\nonumber\\
&= & e^{i\delta{ Q} ({x}_i-{ x}_j) }\times \mbox{slow oscillation part},
\end{eqnarray}
with
\begin{equation}
\delta { Q} ({x}_i-{x}_j) \equiv \pm \pi\langle N_{ij}^h\rangle.
\end{equation}
To leading approximation, $\Pi^h_{ij}$ will then contribute to an {\it additional} oscillation 
factor at wavevector 
\begin{equation}
\delta Q=\pm\pi \delta/a.
\end{equation}
Besides such a singular oscillating part, there is a slow oscillation part in (\ref{ff}) which
only gives rise to an additional decay, determined 
at $J\rightarrow 0^+$ limit as\cite{string1} $\sim 1/ |x_{ij}|^{1/2}$ at 
$|x_{ij}|\equiv|x_i-x_j|\gg 
a$, and thus
\begin{equation}\label{f1}
\langle\Pi^h_{ij}\rangle \sim \frac{e^{i\delta Q {x}_{ij}} } {|x_{ij}|^{1/2}}.
\end{equation}
It then leads to an incommensurate momentum structure at $Q_0+\delta Q=\pm(\pi/a)[1-\delta]
\equiv \pm 2k_f$ in 1D spin-spin correlation function. Including the time $t$ in (\ref{f1}) only changes
the decay factor $|x_{ij}|^{-1/2}$ to $(x_{ij}^2-v_c^2t^2)^{-1/4}$ at long-time 
limit ($v_c$ is the Fermi velocity of holon) without modifying the incommensurate position at $2k_f$. The correct power-law decay of the spin-spin correlation function at $J/t\rightarrow 0^+$ limit is given by:
\begin{equation}
\langle S^+_i(t)S^-_j(0)\rangle\sim \frac{\cos (2k_fx_{ij})}{(x_{ij}^2-v_s^2t^2)^{1/2}
(x_{ij}^2-v_c^2t^2)^{1/4}},
\end{equation}
where $v_s$ is the spinon velocity. Furthermore, the similar phase-string effect 
also gives rise to a Luttinger-liquid type Fermi-surface singularity in the momentum distribution of 
electrons at $k_f$ with an exponent $1/8$.\cite{string1} Even though the $2k_f$ incommensurate 
structure in spin-spin correlations is still connected to the Fermi-momentum 
$k_f$ here, a Fermi-liquid type interpretation in terms of the Kohn anomaly is no 
longer accurate since the electron quasiparticle description is not elementary anymore. A correct picture is that each holon as a topological object 
carries a spin domain-wall with it so that in calculating the large-distance 
spin-spin correlation function one has to count how many holons in between
as each of them contributes a $\pi$ phase-shift as shown in (\ref{phase1d}).

\subsection{Incommensurate structure in 2D}

The spin-charge separation in 2D is slightly different from 1D. While in 
1D spinons and holons can be treated as {\it decoupled} at low-energy and 
long-wave length limit, spinons and holons in 2D, though they still belong to 
independent 
degrees of freedom, will influence each other through some topological 
gauge fields even in long-distance limit. For example, spinons are subjected to 
a lattice gauge field $A_{lm}^h\neq 0$ in 2D in the Hamiltonian\cite{string1} (here $(lm)$
refers to a nearest-neighbor link). As a physical quantity, the spin-spin
correlation
function defined in (\ref{s+-}) should be gauge-invariant, independent of the
gauge choice of $A_{lm}^h$. This requires $\Pi^h_{ij}$ to be able to absorb
the gauge phase arising from the spinon propagator under a gauge transformation: 
$A_{lm}^h\rightarrow A_{lm}^h+ (\theta_l-\theta_m)$  and $b_{l\sigma}\rightarrow
b_{l\sigma}e^{i\sigma\theta_l}$ which leaves the Hamiltonian 
unchanged.\cite{string1}  

In fact, $\Pi^h_{ij}$ itself can be expressed in terms of a gauge field. 
Introduce the following relation
\begin{equation}\label{phase1}
\mbox{Im ln $(z_i-z_l)$}-\mbox{Im ln $(z_{j}-z_l)$}=\int_{\Gamma} d{\bf r}\cdot
\frac{{\bf \hat{z}}\times ({\bf r}-{\bf r}_l)}{|{\bf r}-{\bf r}_l|^2},
\end{equation}
where $\Gamma$ is an arbitrary path connecting $i$ and $j$ without crossing 
the site $l$. In the following we always choose $\Gamma$ as one
of the shortest-paths between $i$ and $j$ without crossing {\it any} lattice site. 
Then ${\Pi}^h_{ij}$ at $t=0$ can be expressed as
\begin{equation}\label{phase}
{\Pi}^h_{ij}=e^{i\left(\Phi^h_i-\Phi^h_j\right)}=e^{i2\int_{\Gamma}d{\bf r}\cdot {\bf {\hat 
A}}^h({\bf r})},
\end{equation}
where
\begin{equation}
{\bf {\hat A}}^h({\bf r})=\frac 1 2 \sum_{l}n_l^h \frac{{\bf \hat{z}}\times ({\bf r}-{\bf r}_l)}{|{\bf r}-{\bf r}_l|^2}.
\end{equation}
Similarly  the lattice gauge field $A_{lm}^h$ in the Hamiltonian\cite{string1} can be also 
expressed by such a vector potential ${\bf {\hat A}}^h({\bf r})$ as follows
\begin{equation}\label{eah}
A^h_{lm}=\int_m^ld{\bf r}\cdot {\bf {\hat A}}^h({\bf r}).
\end{equation}
Then it is straightforward to verify the gauge invariance of the spin-spin
correlation function defined in (\ref{s+-}).

In the mean-field theory, the gauge field $A_{lm}^h$ defined at a
nearest-neighbor link is treated by smearing the holon number distribution 
locally.\cite{string2} It is equivalent to replacing ${\bf {\hat A}}^h({\bf r})$
in (\ref{eah}) by its continuum version ${\bf {A}}^h({\bf r})$ defined by
\begin{equation}
{\bf  A}^h({\bf r})=\frac 1 2 \int d^2{\bf r}'\rho^h({\bf r}')\frac{{\bf \hat{z}}\times ({\bf r}-{\bf r}')}{|{\bf r}-{\bf r}'|^2},
\end{equation}
where the holon density $\rho^h({\bf r}')$ is obtained by
smearing the holon-number distribution on lattice site, $n_l^h$, 
in the continuum space at a scale of $a$. However, if one naively does the
same replacement in $\Pi^h_{ij}$ [(\ref{phase})] for a larger spatial separation
of $i$ and $j$, an important effect similar to the singular phase string effect 
discussed in 1D would be lost.

Generally one may define 
\begin{equation}\label{deltaphi}
\Delta\phi_{\Gamma}\equiv 2\int_{\Gamma}d{\bf r}\cdot \left[\hat{\bf 
A}^h({\bf r})-{\bf A}^h({\bf r})\right],
\end{equation}
and rewrite
\begin{equation}\label{phase2}
\Pi_{ij}^h= e^{i2\int_{\Gamma}d{\bf r}\cdot {\bf { 
A}}^h({\bf r})+i\Delta \phi_{\Gamma}}.
\end{equation}
$\Delta\phi_{\Gamma}$ will then keep track of the aforementioned singular effect
in passing from ${\bf\hat{A}}^h({\bf r})$ to ${\bf{A}}^h({\bf r})$. 

According to (\ref{phase1}), the contribution to $2\int_{\Gamma}d{\bf r}\cdot 
{\hat{\bf A}}^h({\bf r})$ from a holon sitting at site $l$ is given by the 
angle spanned between two straight lines connecting
$l$--$i$ and $l$--$j$ as shown in Fig. 2a. The r.h.s. of (\ref{phase1}) is a 
continuous function of ${\bf r}_l$ as long as ${\bf r}_l$ stays away from the 
path $\Gamma$ connecting $i$ and $j$. So when $l$ is shifted at a scale of $a$ 
in the limit of $|{\bf r}_i-{\bf r}_j|\gg a$, the change of the corresponding
angle is negligibly small. In this case, taking a continuum limit by neglecting the
underlying lattice can be justified. But if ${\bf r}_l$ crosses the branch-cut 
$\Gamma$, then it is easy to see that the angle defined in (\ref{phase1}) has a
$2\pi$ jump. Fig. 2a shows a nearest-neighbor lattice link that is across 
the path $\Gamma$ and the value of (\ref{phase1}) changes from $\sim\pi$ to 
$ -\pi$, namely, by $\sim 2\pi$ for the two lattice sites of the link. It is 
important to note that 
in $\Pi_{ij}^h$ such a $2\pi$ change has no effect and thus a holon sitting at either
lattice site of the nearest-neighbor link across $\Gamma$ has approximately the same 
contribution to $\Pi_{ij}^h$ by a phase shift $\pi$. On the other hand, ${\bf A}^h({\bf r})$
is obtained by taking a continuum limit with holon distribution being `coarse-grained' at scale 
of $a$, and in $2\int_{\Gamma}d{\bf r}\cdot {\bf A}^h({\bf r})$ the contribution 
from those holons near the path $\Gamma$ within the lattice scale of $a$ will be simply cancelled 
out due to the opposite signs across $\Gamma$ as shown in Fig. 2a.  Thus,
we find 
\begin{equation}
e^{i2\int_{\Gamma}d{\bf r}\cdot \hat{\bf  A}^h({\bf r})}\simeq e^{i2\int_{\Gamma}d{\bf r}\cdot {\bf { 
A}}^h({\bf r})\pm i\pi N_h(\Gamma)},
\end{equation}
where $N_h({\Gamma})$ denotes the total number of holons at those
nearest-neighboring links across the path $\Gamma$ shown in Fig. 2b.

So the nonlocal phase shift in 2D is identified by
\begin{equation}\label{shift}  
\Delta\phi_{\Gamma}\simeq\pm \pi  N_h({\Gamma}).
\end{equation}
As shown in Fig. 2b, the path $\Gamma$ is always `covered' by
nearest-neighbor lattice links along the $\hat{y}$ 
axis if $|x_i-x_j|\geq|y_i-y_j|$. Noting that each site the average holon number is $\delta$, one has 
\begin{equation}
\langle N_h({\Gamma})\rangle \simeq 2\delta \frac {|x_i-x_j|}{a}.  
\end{equation}
Similarly, the path $\Gamma$ in Fig. 2b 
can be covered by nearest-neighbor lattice links along $\hat{x}$ axis if
$|x_i-x_j|\leq|y_i-y_j|$, and the 
effective holon number along the path $\Gamma$ is estimated to be
$2\delta|y_i-y_j|/a$. Thus, two kinds of oscillations in real space are 
combined to give
\begin{equation}\label{inc}
e^{i\Delta \phi_{\Gamma}}\sim \frac{1}{2}\left\{\theta_{ij}^x\cos [2\pi\delta/a 
(x_i-x_j)] + \theta_{ij}^y\cos [2\pi\delta/a(y_i-y_j)] \right\},
\end{equation}
where $\theta_{ij}^{x(y)}=1$ if $|x_i-x_j|\geq (<) |y_i-y_j|$ and 
otherwise $\theta_{ij}^{x(y)}=0$. 

Finally, the spin-spin correlation function (\ref{s+-}) is expressed by
\begin{eqnarray}\label{ss}
\left\langle S^{+}_i(t) S^{-}_j(0)\right\rangle&\simeq &
e^{i{\bf Q}_0\cdot ({\bf r}_i-{\bf r}_j)+i\Delta\phi_{\Gamma}}
\left\langle b^{\dagger}_{i\uparrow}(t)b_{i\downarrow}(t)
\left(e^{i2\int_{\Gamma}d{\bf r}\cdot {\bf { 
A}}^h({\bf r})}\right) b^{\dagger}_{j\downarrow}(0)b_{j\uparrow}(0)\right\rangle.
\end{eqnarray}
The singular phase string effect is thus retained in the phase factor 
$e^{i\Delta\phi_{\Gamma}}$ and the gauge-invariant quantity inside $\langle ... \rangle$ on the r.h.s. can be now evaluated by the mean-field approximation. In
obtaining (\ref{ss}) we have made approximations by neglecting the fluctuations 
in $\Delta \phi_{\Gamma}$ itself as well as the temporal dependence of 
$\Pi^h_{ij}(t)$. These effects in 1D lead to
an additional decay of the spin-spin correlation function in spatial and
temporal space but do not change the incommensurate momentum position. 
We expect the same thing to happen in 2D where these effects merely cause 
some extra broadening of magnetic peaks. These broadening effects should
become minimal at low-temperature when a Bose condensation of holons occurs, 
which corresponds to a superconducting condensation\cite{string2} (in the phase
string theory both spinon and holon are {\it bosonic}).

Therefore, similar to 1D, the spatial oscillation factor
$e^{i\Delta\phi_{\Gamma}}$ 
in (\ref{ss}) will lead to incommensurate momentum shifts from ${\bf Q}_0$
by $\delta {\bf Q }= (\pm \bar{\delta}\pi/a, 0)$ and $(0,\pm \bar{\delta}\pi/a)$ 
according to
(\ref{inc}) [It is noted that $\theta_{ij}^{\eta}$ in(\ref{inc}) will
modify the plane-wave oscillation function but will not change the peak 
positions at $\delta{\bf Q}$]. The incommensurability $\bar{\delta}= 
2\delta$ obtained here is in agreement with the experiments discussed in the
beginning. Of course, whether such an incommensurate structure can become 
{\it observable} also depends on the broadening effect introduced by the spinon
propagator [the average on the r.h.s. of (\ref{ss})].
With the singular phase string effect being explicitly sorted out, such a
gauge-invariant quantity can be directly calculated in terms of 
the mean-field theory developed in Ref.\onlinecite{string2}. For the sake of
compactness, the detailed mathematical manipulation is given in the 
Appendix. The dynamic spin susceptibility function $\chi''(i,j;\omega)$ is 
shown in (\ref{ichi+-}) (transverse component) and (\ref{ichizz})  
($z$-component) and is rotationally invariant at the mean-field level. 

In Ref.\onlinecite{string2}, two kinds of mean-field solution
have been found for metallic phase: a uniform phase without 
a Bose condensation of spinons and an inhomogeneous phase with the Bose 
condensation of spinons at low temperature. Mathematically they are controlled 
by the strength of fluctuations in the gauge field $A^h_{ij}$ defined in 
(\ref{ah}). Note that the Bose condensation of
spinons at half-filling corresponds to a long-range AF ordering of spins. In the
metallic phase, the long-range AF order is generally absent. But in the spinon
Bose condensation case, a tendency towards the AF ordering is still present as 
exhibited in a form of inhomogeneity (microscopic phase separation) and 
is argued to occur usually at smaller doping (underdoped regime).
An experimental characterization of those two kinds of metallic phase has been
also proposed based on the distinct energy structure\cite{string2} of the 
local dynamic spin susceptibility function $\chi''_L(\omega)$, which is 
illustrated in the insert of Fig. 3 for the same doping concentration 
$\delta=0.143$ and at $T=0$. 
The dashed curve corresponds to the uniform phase where at low energy there only 
a resonance-like peak at $\omega = 2E_s\sim 0.4 J$ (J is the superexchange 
coupling constant) and a real spin gap below the peak; the solid curve corresponds to the inhomogeneous phase with a spinon Bose condensation where a 
double-peak is shown with a nonzero but suppressed weight at low energy
resembling a pseudo-gap picture. 

We compute the momentum-dependence of the dynamic spin 
susceptibility function corresponding to those two cases with the same 
parameters. Fig. 3 shows $\chi''({\bf q}, \omega)$ as
the Fourier transformation of $\chi''(i,j;\omega)$ given in the Appendix. 
The momentum scan is illustrated by the left top insert in Fig. 3 which is along 
$q_x$-axis at a fixed $q_y=\pi/a$, and four full circles denote the 
incommensurate peak positions at ${\bf  Q}_0+\delta{\bf Q}$. First of all, the dashed curve corresponds to 
aforementioned uniform-phase case at $\omega=2E_s$. It shows no explicit
incommensurate peak splitting because the 
average in (\ref{ss}) contributes to a large broadening 
in momentum space which smears the peak structure composed of four 
incommensurate peaks. The corresponding resonance-like peak around $\omega=2E_s$ 
in the insert of Fig. 3 was used to explain\cite{string2} the $41$ $meV$ peak in 
the optimally doped $YBa_2Cu_3O_7$ compound.\cite{o7} A single 
{it broad} AF peak shown in Fig. 3 is also consistent with the 
neutron-scattering experiments in such a compound.

The incommensurate peaks at ${\bf Q}_0+\delta{\bf Q}$ explicitly show up in the 
solid curve (inhomogeneous phase) in Fig. 3 at small $\omega$. Since there is a 
spinon Bose condensation in this case, the susceptibility function has nonzero 
weight persists over to low energy until $\omega=0$ (the insert of Fig. 3). 
In this case, the spin-spin correlation length at small $\omega$ becomes much 
longer than the uniform-phase case such that the underlying incommensurate structure becomes visible.  Experimentally, the incommensurate 
splitting has been indeed observed at small energy transfer in $LSCO$
compounds\cite{inc1,inc2,inc3,inc4} at the same positions and with the same 
doping-dependent incommensurability $\bar{\delta}$ shown in Fig. 3. Most recently, a similar 
incommensurate structure has been also identified in the underdoped $YBCO$ compounds.\cite{inc5}

\subsection{Reconciliation of the incommensurability with NMR measurements} 

The NMR experiment has served as an another powerful tool in probing the 
magnetic properties in cuprates. A combination of NMR and neutron-scattering
measurements should provide a more complete picture for the nature of AF 
spin-spin correlations in these materials.   

The NMR spin-lattice relaxation rate for nuclear spins at planar $^{17}O$ sites,
$1/^{17}T_1$, can be expressed in terms of the real-space spin-spin 
correlations as follows\cite{mila,shastry,scalapino,mmp}
\begin{equation}\label{t17}
\frac {1} {^{17}T_1}\sim \frac {T} N \sum _{i} \left[G_{ii}+\frac 1 4 
\sum_{j=nn(i)}G_{ij}\right]
\end{equation}
where
\begin{equation}
\left.G_{ij}\equiv \frac{\chi'' (i,j, \omega)}{\omega } \right|_{\omega \rightarrow 0^+} . 
\end{equation}
According to (\ref{t17}), $1/^{17}T_1$ 
only involves real-space spin-spin correlations up to nearest-neighboring sites. 
The spin relaxation rate for planar $^{63}Cu$ nuclear spin can be similarly 
obtained\cite{mila,shastry,scalapino,mmp} which involves spin-spin correlations up to 
next-nearest-neighboring sites. So in the one-component 
theory\cite{mila,shastry,scalapino,mmp} of the NMR in cuprates, only 
short-distance spin-spin correlations are relevant.
  
It has been well known\cite{scalapino,mmp} that a sharp commensurate AF 
component in $\chi''(i,j)$ will 
have strong cancellation between the on-site and nearest-neighboring-site terms 
in (\ref{t17}) due to the staggered factor $(-1)^{i-j}$. So commensurate AF 
correlations, which presumably are responsible to a so-called non-Korringa 
temperature behavior in $1/^{63}T_1$, will have much less significant 
contribution to $1/^{17}T_1$, as has been discussed by many 
authors,\cite{scalapino,mmp} and this is consistent with
experimental measurements where $1/^{17}T_1$ shows a much weaker and quite different behavior usually attributed to the non-AF correlations in $\chi''$.

However, the neutron-scattering measurements indicate AF correlations in 
cuprates should be {\it incommensurate} in nature as discussed in the previous
section. Once the sharp commensurate AF component is replaced by an incommensurate AF structure experimentally, it
was pointed out\cite{wsc} that the non-Korringa 
behavior of $1/T_1$ at $^{63}Cu$ would usually `leak' to the $1/^{17}O$ sites,  
causing a temperature behavior in $1/^{17}T_1$ which is not consistent with experimental
results. This is due to an imperfect cancellation of the AF component of 
$\chi''$ in (\ref{t17}) after the change from a sharp commensurate to an 
incommensurate (or a more broadened) magnetic structure around ${\bf Q}_0$. Indeed, various theories\cite{fermi1,fermi2,fermi3} based on the Fermi surface 
topology in 
which the incommensurability arises from the Kohn anomaly have predicted
a {\it similar} non-Korringa behavior for $1/^{17}T_1$, in contrast to the experimental
results.  In this sense, incommensurate AF correlations seem inconsistent with
the NMR data. In order to resolve this issue, some authors\cite{zha} even have 
proposed to modify the form of hyperfine coupling as an alternative way out.

But such a paradox can be easily reconciled in the phase string description without changing the
hyperfine interactions which are determined based on the local chemical structure of 
cuprates and are consistent with the Knight shift data.\cite{mila,shastry,scalapino,mmp} 
In the present theory, the incommensurate structure or spatial modulation, as represented by
$e^{i\Delta\phi_{\Gamma}}$ in (\ref{ss}),  has been directly attributed to the phase vortex structure whose singular centers are bound to holons. 
The resulting incommensurate oscillation in spin-spin correlation function 
is only meaningful at a scale much larger than the size of the singular center.   
As for a nearest-neighbor lattice link $(ij)$, $\Pi^h_{ij}$ will simply reduce to
\begin{equation}
\Pi^h_{ij}=e^{i\Phi^h_i}e^{-i\Phi^h_j}=e^{iA^h_{ij}},
\end{equation}
where $A^h_{ij}$ is the lattice gauge field given in (\ref{eah}). In the mean-field approximation $A^h_{ij}$ is replaced by $\int^i_jd{\bf r}\cdot 
{\bf A}^h({\bf r})$ as discussed before, without causing any additional singular 
phase shift, namely, $\Delta\phi_{\Gamma}=0$. It is 
because there can be no holon sitting between two nearest-neighboring spins.
Trivially, one also has $\Pi^h_{ii}=1$. Thus, $1/^{17}T_1$ in (\ref{t17})
will simply not be able to `see' any incommensurate phase shift, even though 
such a incommensurate spatial modulation does exist in large-distance correlations as
given by (\ref{inc}). The latter can be picked up by the neutron-scattering measurement
as discussed in previous section. 

Fig. 4 shows an example of a non-Korringa behavior of $1/^{63}T_1$ vs. a 
Korringa-like behavior of $1/^{17}T_1$ based on the dynamic spin susceptibility function 
determined in the Appendix where the uniform (optimal-doping) mean-field solution\cite{string2} is used at $\delta=0.143$. The distinct temperature-dependence and the relative 
size of $1/^{63}T_1$ and $1/^{17}T_1$ here are in accord with the typical experimental data
of cuprates. $1/^{63}T_1T$ exhibits a non-Korringa $1/T$ behavior at high temperature while
$1/^{17}T_1T$  remains roughly a constant (Korringa behavior). Note that the decrease of the relaxation
rates at low-$T$ is due to the spin gap $\sim E_s$ which also decides the superconducting
transition temperature $T_c$ around the same scale.\cite{string2} One thing 
we need to emphasize here is that even though in the uniform-phase case the
momentum structure in Fig.3 looks like a commensurate peak, it actually is
composed of four incommensurate peaks with stronger broadening, and thus the 
`leakage' effect would be expected even stronger in this case. In fact if we 
assume the incommensurate spatial modulation (\ref{inc}) to persist into 
short-distance 
$\sim a$, the non-Korringa temperature dependence will then immediately show up 
in $1/^{17}T_1$ precisely as pointed out in Ref.\onlinecite{wsc}. Therefore,
the origin of the incommensurability from the nonlocal vortex structure of 
(\ref{phih}), i.e., the phase string effect, is indeed crucial to resolve this 
issue.

\section{DISCUSSIONS}

In the present paper, we have related the magnetic incommensurability to the intrinsic 
properties of the doped Mott insulator. In the Mott insulator, 
the spins are the only degrees of freedom not frozen by the strong on-site Coulomb interaction.
Holes doped into such a Mott insulator always introduces a phase frustration 
(phase string effect) on the spin background by violating the Marshall sign 
rule.\cite{string,string1} Such a phase string effect then contributes a unique 
spatial modulation to the spin-spin correlations as demonstrated in this paper,
which is more or less independent of 
energy and temperature but linearly depends on the doping concentration. In 
momentum space it leads to an explicit incommensurate splitting of the AF peak 
in the dynamic spin susceptibility when the spin correlation length is 
sufficiently long, in agreement with the neutron-scattering measurements
of cuprates in both the peak positions and doping dependence. Furthermore, due to the phase string origin of this incommensurability, it has been shown that the NMR spin
relaxation rates should not be able to pick up the incommensurability AF
correlations at length scales comparable to the lattice constant, and thus 
conflicting implications of neutron and NMR data can be reconciled within the 
present framework. Conversely, one may also conclude that the magnetic 
incommensurability observed in cuprate superconductors provides a direct 
experimental evidence for the phase string effect in the doped Mott insulator.       
  
We conclude this paper by making several remarks on some related issues. First
of all, we point out that the present magnetic incommensurability is a unique
property of metallic phase, which should disappear in the localized regime.
Recall that the phase string effect leads to a vortex phase $e^{i\Phi^h_i}$ emerging in spin 
operator (\ref{s+}) which is the reason for the incommensurate spatial modulation. 
It has been previously discussed\cite{string2} that the same phase $e^{i\Phi^h_i}$ also causes
the disappearance of the AF long range order. It was argued that only in 
the hole-localized regime (insulating phase) where the phase string effect is no longer effective,
the phase $e^{i\Phi^h_i}$ can be compensated by the vortex phase carried by spinons
which become vortices in the insulating phase, and the AF long range order may
be recovered. By the same token, the incommensurability induced by 
$e^{i\Phi^h_i}$ should be also gone in this localized regime 
at small doping. This is in accord with the experiments where 
only a commensurate AF peak has been observed in the insulating phase.\cite{inc4}     

Secondly, as briefly mentioned in Sec. IIA, even though $2k_f$ incommensurate
oscillation of 1D spin-spin correlation function may be viewed as the Kohn 
anomaly of the Fermi surface effect, it is not very meaningful because single
electron excitations are no longer {\it elementary} and the spin-spin correlation
function cannot be reasonably approximated by a `bubble' diagram composed of
the single-electron Green's function without including high order
corrections. Similarly the 2D magnetic incommensurability in the phase
string formalism does not necessarily exclude the relevance of the Fermi surface
topology. Instead it implies that the latter explanation is no longer accurate
as holon and spinon become new elementary excitations. Especially,
the interpretation of the doping-dependent incommensurability and the 
reconciliation of neutron and NMR measurements can only be naturally realized in
this phase string description of the doped-Mott-insulator.
 
Finally, we emphasize that the 
incommensurability discussed in the present paper is due to the topology
associated with each individual holon. It has nothing directly to do with the phase
separation or charge inhomogeneity at a larger length scale involving many holes as discussed 
in literature (like the stripe phase\cite{stripe}), including the
inhomogeneous (underdoping) metallic regime studied within the phase string 
formalism.\cite{string2} Nevertheless, there is still 
a fundamental connection: all of these may be attributed to the competition 
between the kinetic energy of holes and superexchange energy of spins and the 
resulting tendency towards phase separation. The vorticity surrounding each 
holon may be regarded as a microscopic version of phase separation against a 
uniform spiral phase.\cite{spiral}        
 
\acknowledgments

The authors acknowledge useful conversations with T. K. Lee, J. X. Li, C. 
Y. Mou, R. E. Walstedt, and X. G. Wen. The authors would also like to 
thank the hospitality of the National Center for Theoretical Science in Taiwan 
where part of the present work was done during their visit. The 
present work is supported, in part, by the Texas ARP program 
No. 3652707 and a grant from the Robert A. Welch foundation, and the State of 
Texas through the Texas Center for Superconductivity at University of Houston. 

\appendix
\section*{DYNAMIC SPIN SUSCEPTIBILITY FUNCTION}

\subsection{Mean-Field Theory}

The $t-J$ model in the phase string formalism can be written in two terms in the 
mean-field approximation:\cite{string2}
$H_s+H_h$. $H_s$ governs the spinon degrees of freedom and $H_h$ determines the
holon degrees of freedom. In the present paper, we mainly focus on the spin 
magnetic properties decided by $H_s$: 
\begin{equation}\label{hs}
H_s=\sum_{m\sigma} E_m \gamma_{m\sigma}^{\dagger}\gamma_{m\sigma} + \mbox{constant},
\end{equation}
where $\gamma_{m\sigma}^{\dagger}$ is the creation operator of bosonic spinon
elementary excitations. The spinon excitation spectrum $E_m$ in this mean-field
theory is given by
\begin{equation}\label{espinon}
E_m=\sqrt{\lambda_m^2-\xi_m^2},
\end{equation}
in which, $\lambda_m$ is defined by
\begin{equation}\label{lambda2} 
\lambda_m=\lambda-\alpha|{\xi}_m|,
\end{equation}
where $\lambda$ is the Lagrangian multiplier to be determined by the condition of total spin number equal to $N(1-\delta)$ ($N$ is the total number of lattice 
sites.) The coefficient $\alpha$
is a ratio of two energy scales: $\alpha=J_h/J_s$, in which $J_h=\delta t'$
describes an effective hopping effect on spinon degrees of freedom with $t'\sim
J$ and $J_s=(J/2)\Delta^s$ with $\Delta^s$ being the bosonic RVB order parameter 
characterizing the present mean-field state. The Lagrangian multiplier $\lambda$
and the RVB order parameter $\Delta^s$ are determined by the following 
mean-field equations, respectively,
\begin{equation}\label{lambda}
2-\delta=\frac 1 N \sum_{m} \frac {\lambda_m}{E_m}\coth \frac{\beta E_m}{2},
\end{equation} 
and
\begin{eqnarray}\label{ds}
\Delta^s=\frac{1}{4N}\sum_{m}\frac{{\xi}_m^2}{J_s{E}_m}\coth \frac { {\beta}{E}_m}{2}.
\end{eqnarray}

The spectrum $\xi_m$
will be modified by doping effect through a gauge field $A_{ij}^h$ as 
the eigenvalue of the following equation\cite{string2}
\begin{equation}\label{ew}
\xi_m w_{m\sigma}(i)= -J_s \sum_{j=nn(i)}e^{i\sigma A_{ij}^h}w_{m\sigma}(j).
\end{equation}
Here $A^h_{ij}$ (with $j=nn(i)$) is originally defined by the following gauge 
invariant condition
\begin{equation}\label{ah}
\sum_CA_{ij}^h=\pi N_C^h,
\end{equation}
where $C$ is an arbitrary counterclockwise closed path and $N_C^h$ is the 
number of holes enclosed by it. In the present mean-field theory, the dynamic
effect of $A_{ij}^h$ is replaced by a randomness as it comes from an independent
degrees of freedom (holons). $N_C^h$ may be rewritten as $\bar{N}_C^h
+\delta N_C^h$ with fluctuating $\delta N_C^h$. Correspondingly 
$A_{ij}^h$ describes a uniform fictitious magnetic field with a random flux
fluctuation on the top of it. Two kinds of mean-field solutions due to different strengths
of the fluctutations are found,\cite{string2} which are distinguished by whether
there is a Bose condensation of spinons or not at low temperature. The former case is
called inhomogeneous metallic phase and the latter is called 
uniform metallic phase in Ref.\onlinecite{string2}. 

Finally, the Bogoliubov transformation relates $\gamma$-operator to 
the spinon operator $b_{i\sigma}$ is given as follows:
\begin{equation}\label{bogo}
b_{i\sigma}=\sum_m\left(u_m\gamma_{m\sigma}-v_m \gamma^{\dagger}_{m-
\sigma}\right)e^{i\sigma\chi_m}w_{m\sigma}(i),
\end{equation}
where $u_m=1/\sqrt{2}(\lambda_m/E_m+1)^{1/2}$ and 
$u_m=1/\sqrt{2}(\lambda_m/E_m-1)^{1/2} sgn (\xi_m)$. The phase factor $e^{i\sigma \chi_m}$ is determined up to a change
\begin{equation}\label{wphase}
e^{i\sigma\chi_m}\rightarrow -sgn(\xi_m)\times e^{i\sigma\chi_m}
\end{equation}
each time when a holon changes sublattices.\cite{string2}

\subsection{Dynamic spin susceptibility}

Let us first consider the transverse component defined in Matsubara 
representation as follows 
\begin{equation}\label{chi+-t}
\chi_{+-}(i,j; i\omega_n)=\int_0^{\beta}d\tau \ \ e^{i\omega_n\tau}
\langle T_{\tau} S^{+}_i(\tau) S^{-}_j(0)\rangle.
\end{equation}
Here $\omega_n=\frac{2\pi n}{\beta}$ and
\begin{equation}\label{+-}
\left\langle T_{\tau} S^{+}_i(\tau) S^{-}_j(0)\right\rangle  = (-1)^{i-j}
\left\langle T_{\tau} b^{\dagger}_{i\uparrow}(\tau)b_{i\downarrow}(\tau)
\Pi^h_{ij}(\tau)b^{\dagger}_{j\downarrow}(0)b_{j\uparrow}(0)\right\rangle.
\end{equation}
According to discussions in Sec. IIB, the r.h.s. of (\ref{+-}) can be expressed
by
\begin{equation}
(-1)^{i-j} \left\langle T_{\tau} b^{\dagger}_{i\uparrow}(\tau)b_{i\downarrow}(\tau)
b^{\dagger}_{j\downarrow}(0) b_{j\uparrow}(0) e^{i2\int_{\Gamma}d{\bf r}\cdot {\bf A}^h({\bf r})+ i\Delta\phi_{\Gamma}}\right\rangle.
\end{equation}
In the mean-field approximation one has
\begin{eqnarray}\label{average+-}
\left\langle T_{\tau} b^{\dagger}_{i\uparrow}(\tau)b_{i\downarrow}(\tau)
b^{\dagger}_{j\downarrow}(0) b_{j\uparrow}(0)\right\rangle_s =
\left\langle T_{\tau} b^{\dagger}_{i\uparrow}(\tau)b^{\dagger}_{j\downarrow}(0)
\left\rangle_s \right\langle T_{\tau} b_{j\uparrow}(0) b_{i\downarrow}(\tau) \right\rangle_s\nonumber\\
\ \ \ \ \ \ \ +\left\langle T_{\tau} b^{\dagger}_{i\uparrow}(\tau) b_{j\uparrow}(0) \right\rangle_s\left\langle 
T_{\tau}b_{i\downarrow}(\tau)b^{\dagger}_{j\downarrow}(0) \right\rangle_s,
\end{eqnarray}
where $\langle ... \rangle_s$ denotes the average over the spinon degrees of 
freedom.
According to the mean-field theory outlined in the above section, one has
\begin{eqnarray}
b^{\dagger}_{i\sigma}(\tau)& = & e^{H_s\tau}b_{i\sigma}^{\dagger} e^{-
H_s\tau}\nonumber\\ &=&\sum_m\left(u_m\gamma_{m\sigma}^{\dagger} e^{E_m\tau}-v_m 
\gamma_{m-\sigma}e^{-E_m\tau}\right){ w}^*_{m\sigma}(i)e^{-
i\sigma\chi_m}.
\end{eqnarray}
Then by noting 
${w}^*_{m-\sigma}={w}_{m\sigma}$ one finds 
\begin{eqnarray}
\lefteqn{\left\langle T_{\tau} b^{\dagger}_{i\sigma}(\tau) b^{\dagger}_{j-
\sigma}(0) \right\rangle_s }\nonumber\\
& & =-\sum_m u_m v_m \left({w}^*_{m\sigma}(i){w}_{m\sigma}(j) \right) e^{-i\sigma 
\Delta \chi_m}\left(e^{E_m\tau}\left\langle
\gamma^{\dagger}_{m\sigma}\gamma_{m\sigma}\right\rangle_s + e^{-
E_m\tau}\left\langle\gamma_{m-\sigma}\gamma^{\dagger}_{m-\sigma}\right\rangle_s\right),
\end{eqnarray}
at $\tau>0$ where $e^{-i\sigma\Delta\chi_m}$ is introduced to incorporate the phase shift 
effect (\ref{wphase}). $e^{-i\sigma\Delta\chi_m}$ may be determined based on 
how many exchanges between the spinon and background holons. The spinon will
acquire a phase shift $\pm\pi$ for $\xi_m>0$ due to $-sgn(\xi_m)=-1$ each time 
it exchanges positions with a holon. Thus, if $N_{ij}^h$ denotes 
the number of holons encountered and exchanged with the 
involved spinon on the shortest-path $\Gamma$ connecting $i$ and $j$, the total 
phase shift acquired by the spinon at $\xi_m>0$ is $\pm \pi N_{ij}^h\simeq 
\Delta\phi_{\Gamma}$. Then we may write down 
\begin{equation}\label{wp2}
e^{-i\sigma\Delta\chi_m}=\left\{\begin{array}{ll}
1, & \mbox{if $\xi_m<0$}\\
(-1)^{N^h_{ij}}, & \mbox{if $\xi_m>0$}
\end{array}\right.\end{equation}
It is noted that for each $m$ with $\xi_m<0$, one always
can find a state $\bar{m}$ with $\xi_{\bar{m}}=-\xi_m>0$ with a wavefunction
\begin{equation}\label{wp3}
w_{\bar{m}\sigma}(i)=(-1)^iw_{m\sigma}(i)
\end{equation}
according to (\ref{ew}). Then in terms of (\ref{wp2}) and (\ref{wp3})
we finally have (at $\tau>0$)
\begin{eqnarray}
\lefteqn{\left\langle T_{\tau} b^{\dagger}_{i\sigma}(\tau) b^{\dagger}_{j-\sigma}(0) \right\rangle_s
=\left((-1)^{i-j+N^h_{ij}}-1 \right)}\nonumber\\
& & \times \mathop{{\sum}'}_m u_m v_m \left({w}^*_{m\sigma}(i){w}_{m\sigma}(j) \right)\left[n(E_m) e^{E_m\tau} + (1+n(E_m))e^{-E_m\tau}\right],
\end{eqnarray}
where the summation $\sum'_m$ only runs over those states with $\xi_m<0$.
Similarly, one can also determine
\begin{eqnarray}
\left\langle T_{\tau} b^{\dagger}_{i\sigma}(\tau) b_{j\sigma}(0) \right\rangle_s
=\left((-1)^{i-j+N^h_{ij}}+1\right)\mathop{{\sum}'}_m \left({w}^*_{m\sigma}(i){w}_{m\sigma}(j) \right)\nonumber\\
\ \ \ \ \times \left[u_m^2 n(E_m) e^{E_m\tau}  + v_m^2 (1+n(E_m))e^{-E_m\tau}\right],
\end{eqnarray}
as well as other averages in (\ref{average+-}).

Then after integrate out $\tau$ in (\ref{chi+-t}) and identify $(-1)^{N^h_{ij}}$ with $e^{i\Delta\phi_{\Gamma}}$ in operator form, one obtains
\begin{equation}
\chi_{+-}(i,j; i\omega_n)
\simeq \chi_{+-}^{(-)}(i,j; i\omega_n)+ (-1)^{i-j}e^{i\Delta\phi_{\Gamma}}\cdot \chi_{+-}^{(+)}(i,j; i\omega_n),
\end{equation}
where 
\begin{eqnarray}
\chi_{+-}^{(\pm)}(i,j;i\omega_n)= \mathop{{\sum}'}_{mm'}K_{mm'}^{+-}(i,j)\left[(p_{mm'}^{\pm})^2 \frac {n(E_{m'})-n(E_{m})}{i\omega_n+E_m-E_{m'}}\right. \nonumber\\
\times (l_{mm'}^{\pm})^2 
\left(1+n(E_m)+n(E_{m'})\right)\left.\frac{1}{2}\left(\frac 1 
{i\omega_n+E_m+E_{m'}}-\frac 1 {i\omega_n-E_m-E_{m'}}\right)\right], 
\end{eqnarray}
with
\begin{eqnarray}
K_{mm'}^{+-}(i,j)&=&2\left[e^{i \int_{\Gamma}d{\bf r}\cdot {\bf 
A}^h}w^*_{m\uparrow}(i)w_{m\uparrow}(j)\right]\left[e^{i\int_{\Gamma}d{\bf 
r}\cdot {\bf A}^h}w^*_{m\downarrow}(j)w_{m\downarrow}(i)\right]\nonumber\\ 
&=&\sum_{\sigma}\left[e^{i\sigma \int_{\Gamma}d{\bf r}\cdot {\bf 
A}^h}w^*_{m\sigma}(i)w_{m\sigma}(j)\right]\left[e^{i\sigma \int_{\Gamma}d{\bf 
}\cdot {\bf A}^h}w^*_{m-\sigma}(j)w_{m-\sigma}(i)\right].
\end{eqnarray}
In obtaining the last line, the symmetry $w^*_{m\sigma}=w_{m-
\sigma}$ has been used. The coherent factors, 
$p^{\pm}_{mm'}$ and $l^{\pm}_{mm'}$ are defined by 
\begin{eqnarray}
p^{\pm}_{mm'}& = & u_mu_{m'}\pm v_mv_{m'}, \nonumber\\
l^{\pm}_{mm'}& = & u_mv_{m'}\pm v_mu_{m'}.
\end{eqnarray}
Finally, the dynamic spin susceptibility function $\chi_{+-}''(i,j; \omega)$ can 
be obtained as the imaginary part of $\chi_{+-}$ after an analytic  
continuation $i\omega_n\rightarrow \omega +i 0^+$ is made: 
\begin{equation}\label{ichi+-}
\chi''_{+-}(i,j;\omega)=\Phi_{+-}^{(-)}(i,j; \omega) + (-1)^{i-j} 
e^{i\Delta\phi_{\Gamma}}\cdot \Phi_{+-}^{(+)}(i,j; \omega),
\end{equation}
where 
\begin{eqnarray}\label{phi+-}
\lefteqn{\Phi_{+-}^{(\pm)}(i,j;\omega) =\frac{\pi}{2}\mathop{{\sum}'}_{mm'}K^{+-
}_{mm'}(i,j)\left[\left(1+n(E_m)+n(E_{m'})\right) (l_{mm'}^{\pm})^2 sgn 
(\omega)\right.} 
\nonumber\\
&\left.\cdot\delta(|\omega|-E_m-E_{m'})+ 2 \left(n(E_m)-
n(E_{m'})\right)(p_{mm'}^{\pm})^2\delta (\omega+E_m-E_{m'})\right].
\end{eqnarray}

The longitudinal spin susceptibility function can be similarly obtained:
\begin{eqnarray}\label{chizz}
\chi_{zz}(i,j; i\omega_n)&\equiv & \int_0^{\beta}d\tau e^{i\omega_n}\left\langle
T_{\tau}S^z_i(\tau)S_j^z(0)\right\rangle\nonumber\\
&= & \chi_{zz}^{(-)}(i,j; i\omega_n)+ (-1)^{i-j} 
e^{i\Delta\phi_{\Gamma}}\cdot\chi_{zz}^{(+)}(i,j; i\omega_n),
\end{eqnarray}
where $S^z_i=\sum_{\sigma}\sigma b^{\dagger}_{i\sigma}b_{i\sigma}$ is used and 
\begin{eqnarray}
{\chi_{zz}^{(\pm)}(i,j;i\omega_n)=\frac 1 2 \mathop{{\sum}'}_{mm'} K^{zz}_{mm'}(i,j)
\left[(p_{mm'}^{\pm})^2 \frac {n(E_{m'})-n(E_{m})}{i\omega_n+E_m-E_{m'}}\right. }\nonumber\\
\left. +(l_{mm'}^{\pm})^2 \left(1+n(E_m)+n(E_{m'})\right) \frac 1 2 \left(\frac 
1 {i\omega_n+E_m+E_{m'}}-\frac 1 {i\omega_n-E_m-E_{m'}}\right)\right] 
\end{eqnarray} 
with 
\begin{equation}\label{kzz}
K^{zz}_{mm'}(i,j)\equiv \sum_{\sigma}\left(e^{i\sigma\int_{\Gamma}d{\bf r}\cdot 
{\bf A}^h }w^*_{m\sigma}(i)w_{m\sigma}(j)\right)\left(e^{-
i\sigma\int_{\Gamma}d{\bf r}\cdot {\bf A}^h }w^*_{m'\sigma}(j)w_{m'\sigma}(i)\right),
\end{equation}
where the phase factors $e^{\pm i\sigma \int_{\Gamma}d{\bf r}\cdot {\bf A}^h}$
are introduced to explicitly show the gauge invariance of $K^{zz}_{mm'}$. 
Generally, one expects the gauge-invariant quantity 
$e^{i\sigma\int_{\Gamma}d{\bf r}\cdot{\bf A}^h}w^*_{m\sigma}(i)w_{m\sigma}(j)$ to be independent 
of $\sigma$ or to be real, which in turn means $K_{mm'}^{zz}=K^{+-}_{mm'}$. 
Consequently, the rotational invariance of spin-spin correlations is retained. 
The corresponding dynamic spin susceptibility function is given by
\begin{equation}\label{ichizz}
\chi''_{zz}(i,j;\omega)=\Phi_{zz}^{(-)}(i,j; \omega) + (-1)^{i-j} 
e^{i\Delta\phi_{\Gamma}} \Phi_{zz}^{(+)}(i,j; \omega),
\end{equation}
where 
\begin{eqnarray}\label{phizz}
\lefteqn{\Phi_{zz}^{(\pm)}(i,j;\omega) =\frac{\pi}{2}\mathop{{\sum}'}_{mm'}K^{zz
}_{mm'}(i,j)\left[\left(1+n(E_m)+n(E_{m'})\right) (l_{mm'}^{\pm})^2 sgn 
(\omega)\right.} 
\nonumber\\
&\left.\cdot\delta(|\omega|-E_m-E_{m'})+ 2 \left(n(E_m)-
n(E_{m'})\right)(p_{mm'}^{\pm})^2\delta (\omega+E_m-E_{m'})\right].
\end{eqnarray}

\figure{Fig. 1 The topology of a holon due to the phase string: any 
closed loop around it will cut through the phase string once or an odd
number of times. \label{fig1}} 

\figure{Fig. 2a The illustration of the angle defined by (\ref{phase1}) at
site $l$. Such an angle jumps approximately $2\pi$ at 
a nearest-neighbor lattice link (bold one) across the path $\Gamma$ connecting 
sites $i$ and $j$ in the limit $|{\bf r}_i-{\bf r}_j|\gg a$.\label{fig2a}}

\figure{Fig. 2b The path $\Gamma$ connecting $i$ and $j$ is covered by the 
nearest-neighbor links along the $\hat{y}$-axis (bold links) in the case that
$|x_i-x_j|\geq |y_i-y_j|$. \label{fig2b}}

\figure{Fig. 3 Dynamic spin susceptibility function $\chi''({\bf q}, 
\omega)$ versus $q_x$ (with a fixed $q_y$ at $\pi/a$) at $\delta=0.143$. 
Two cases: the dashed curve corresponds to the uniform phase at $\omega= 
2E_s\sim 0.4 J$ where $\chi''_L(\omega)$ is peaked (the insert); the 
solid curve corresponds to the inhomogeneous phase at $\omega\sim 0$. The
$\omega$-dependence of $\chi''_L(\omega)$ in the insert is from Ref. \onlinecite{string2}.
The positions of the incommensurability is illustrated by four full 
circles at the left top. Note that the maxima of $\chi''$ in two cases are 
scaled to the same value for comparison.\label{fig4}}

\figure{Fig. 4 The contrast of the non-Korringa and Korringa behavior of 
the NMR spin relaxation rates for planar copper and oxygen nuclear 
spins for the uniform phase case at $\delta=0.143$. 


\begin{references}
\bibitem{inc1} G. Shirane, R. J. Birgneau, Y. Endoh, P. Gehring, M. A. 
Kastner, K. Kitazawa, H. Kojima, I. Tanaka, T. R. Thurston, and K. Yamada, Phys.
Rev. Lett. {\bf 63}, 330 (1989).
\bibitem{inc2} S-W. Cheong, G. Aeppli, T. E. Mason, H. Mook, S. M. Hayden, P. C.
Canfield, Z. Fisk, K. N. Clausen, and J. L. Martinez, Phys. Rev. Lett. {\bf 67}, 
1791 (1991).
\bibitem{inc3} T. E. Mason, G. Aeppli, and H. A. Mook, Phys. Rev. Lett. {\bf 68}, 1414 (1992).
\bibitem{inc4} K. Yamada, C. H. Lee, K. Kurahashi, J. Wada, S. Wakimoto, S. 
Ueki, H. Kimura, Y. Endoh, S. Hosoya, G. Shirane, R. J. Birgeneau, M. Greven, M. 
A. Kastner, and Y. J. Kim, Phys. Rev. B{\bf 57}, 6165 (1998).
\bibitem{inc5} H. A. Mook, P. Dai, R. D. Hunt, and F. Do\u{g}an, 
cond-mat/9712326.
\bibitem{fermi1} Q. Si, Y. Zha, K. Levin, and J. P. Lu, Phys. Rev. B{\bf 47}, 
9055 (1993).
\bibitem{fermi2} P. B. Littlewood, J. Zaanen, G. Aeppli, and H. Monien,
Phys. Rev. B{\bf 48}, 487 (1993). 
\bibitem{fermi3} T. Tanamoto, H. Kohno, and H. Fukuyama, J. Phys. Soc. Jpn., {\bf 62}, 717 (1993); {\it ibid}, {\bf 63}, 2739 (1994).
\bibitem{anderson} P. W. Anderson, Science {\bf 235}, 1196 (1987); P. W. 
Anderson, {\it The Theory of Superconductivity in the High $T_c$ Cuprates} 
(Princeton Univ. Press, Princeton, 1997).
\bibitem{wsc} R. E. Walstedt, B. S. Shastry, and S-W. Cheong, Phys. Rev. Lett. {\bf 72}, 3610 (1994), and references therein. 
\bibitem{ins} S. Chakravarty, B. I. Halperin, and D. R. Nelson, Phys. Rev.  Lett. {\bf 60}, 1057 (1988).
\bibitem{ins1} see, A. V. Chubukov, S. Sachdev, and J. Ye, Phys. Rev. B{\bf 49}, 11919 (1994) and the references therein.
\bibitem{string} D. N. Sheng, Y. C. Chen, and Z. Y. Weng, Phys. Rev. Lett. {\bf 
77}, 5102 (1996).
\bibitem{string1} Z. Y. Weng, D. N. Sheng, Y. C. Chen, and C. S. Ting, Phys. 
Rev. B{\bf 55}, 3894 (1997).
\bibitem{rvb} S. Liang, B. Doucot, and P. W. Anderson, Phys. Rev. Lett. {\bf 
61}, 365 (1988).
\bibitem{aa} D.P. Arovas and A. Auerbach, Phys. Rev. B{\bf 38}, 316 (1988);
A. Auerbach, Phys. Rev. Lett. {\bf 61}, 617 (1988). 
\bibitem{string2} Z. Y. Weng, D. N. Sheng, and C. S. Ting, Phys. Rev. Lett.
{\bf 80}, 5401 (1998); Z. Y. Weng, D. N. Sheng, and C. S. Ting, preprint.
\bibitem{o7} M. A. Mook, Yethiraj, G. Aeppli, T. E. Mason, and T. Armstrong,
Phys. Rev. Lett. {\bf 70}, 3490 (1993); H. F. Fong, B. Keimer, P. W. Anderson, D. Renzik, F. Do\u{g}an, and I. A. Aksay, Phys. Rev. Lett. {\bf 75}, 316 (1995) 
\bibitem{mila} F. Mila and T.M. Rice, Physica C{\bf 17}, 561 (1989).
\bibitem{shastry} B.S. Shastry, Phys. Rev. Lett. {\bf 63}, 1288 (1989).
\bibitem{scalapino} N. Bulut, D. Hone, D. J. Scalapino, and N. E. 
Bickers, Phys. Rev. Lett. {\bf 64}, 2723 (1990).
\bibitem{mmp} A.J. Millis, H. Monien, and D. Pines, Phys. Rev. B{\bf 
42}, 167 (1990).
\bibitem{zha} Y. Zha, V. Barzykin, and D. Pines, Phys. Rev. B{\bf 54}, 7561 
(1996). 
\bibitem{stripe} J. M. Tranquanda, {\it et al.}, Nature (London) {\bf 375}, 561 
(1995); J. M. Tranquanda {\it et al.}, Phys. Rev. B{\bf 54}, 7489 (1996); J. M. Tranquanda {\it et al.}, Phys. Rev. Lett. {\bf 78}, 338 (1997).
\bibitem{spiral} B. I. Shraiman and E. D. Siggia, Phys. Rev. Lett. 
{\bf 62}, 1564 (1989); {\bf 61}, 467 (1988);
C. Jayaprakash {\it et al.}, Phys. Rev. B {\bf 40}, 2610 
(1989); D. Yoshioka, J. Phys. Soc. Jpn., {\bf 58}, 1516 (1989); 
C. L. Kane {\it et al.}, Phys. Rev. B{\bf 41}, 2653 (1990);
Z. Y. Weng, Phys. Rev. Lett. {\bf 66}, 2156 (1991).

\end{references}
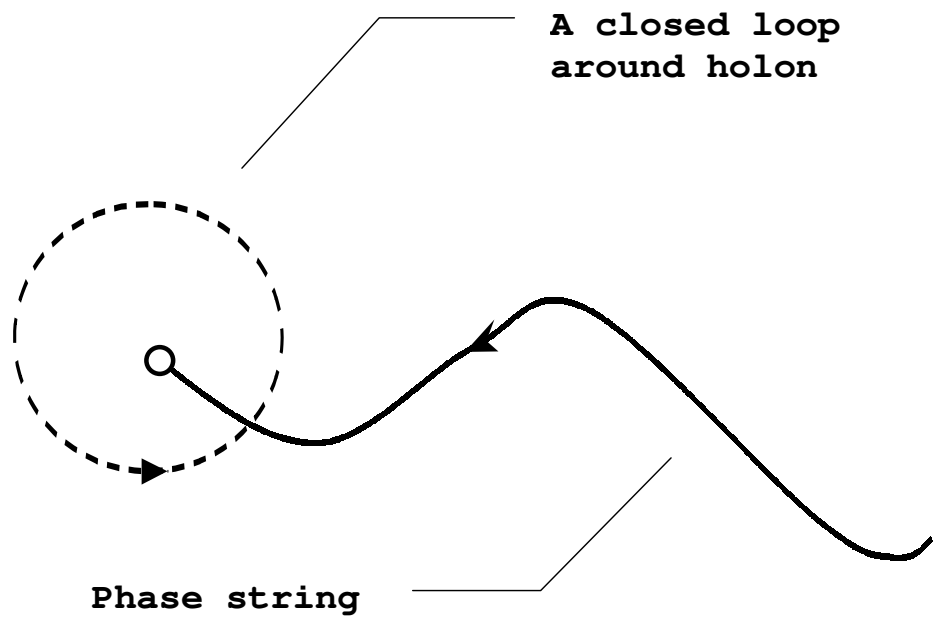
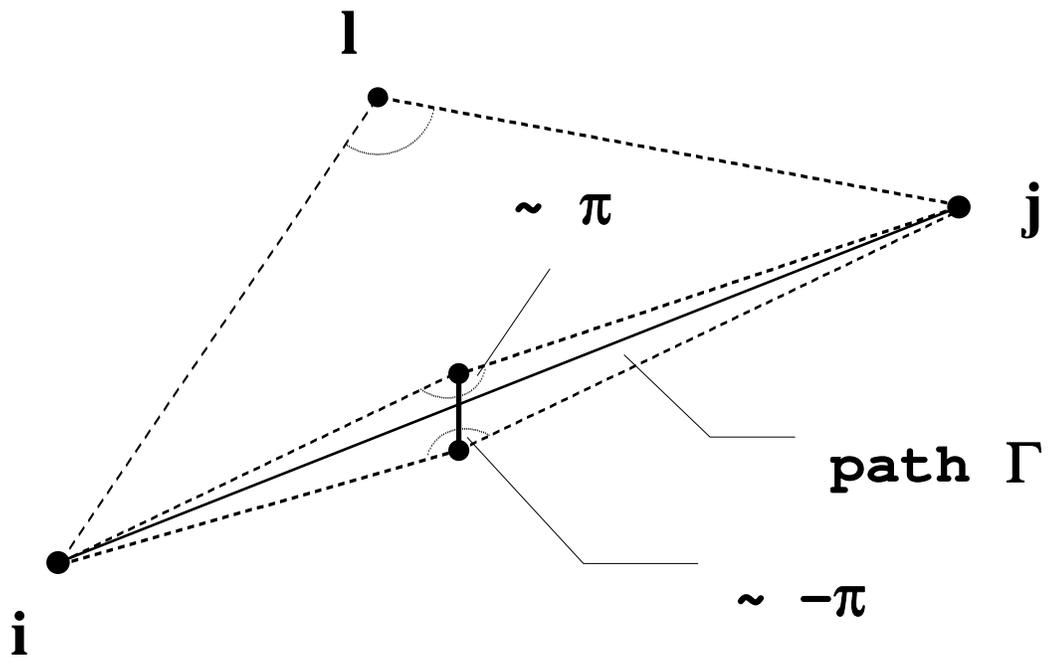
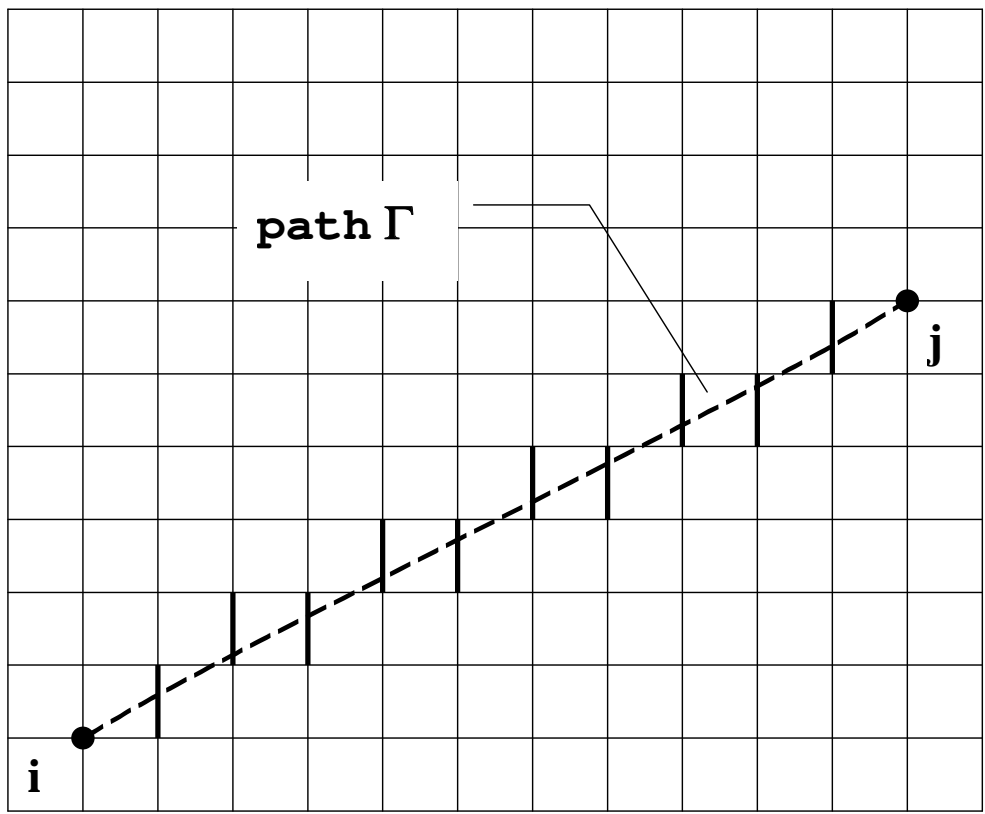
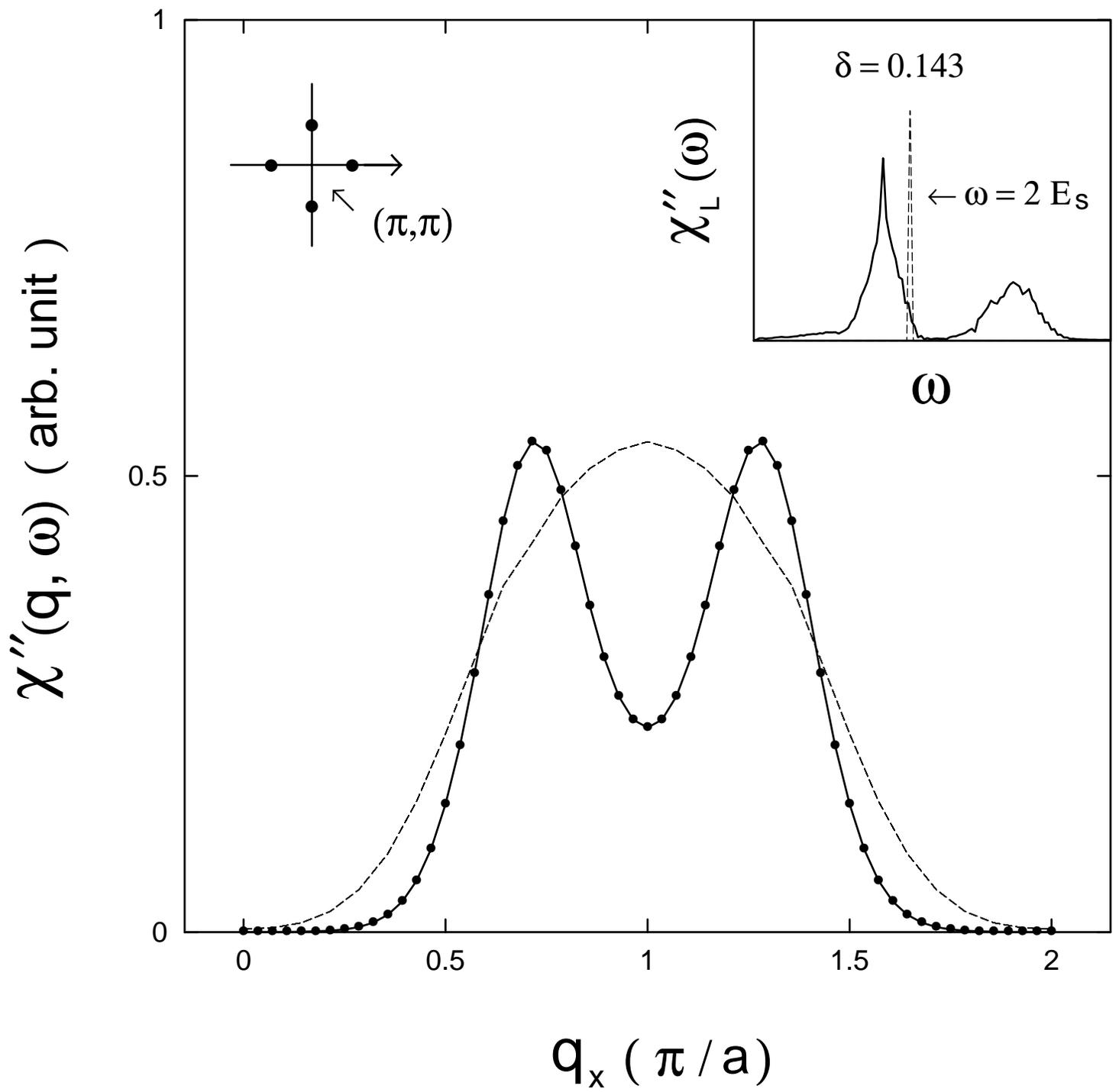
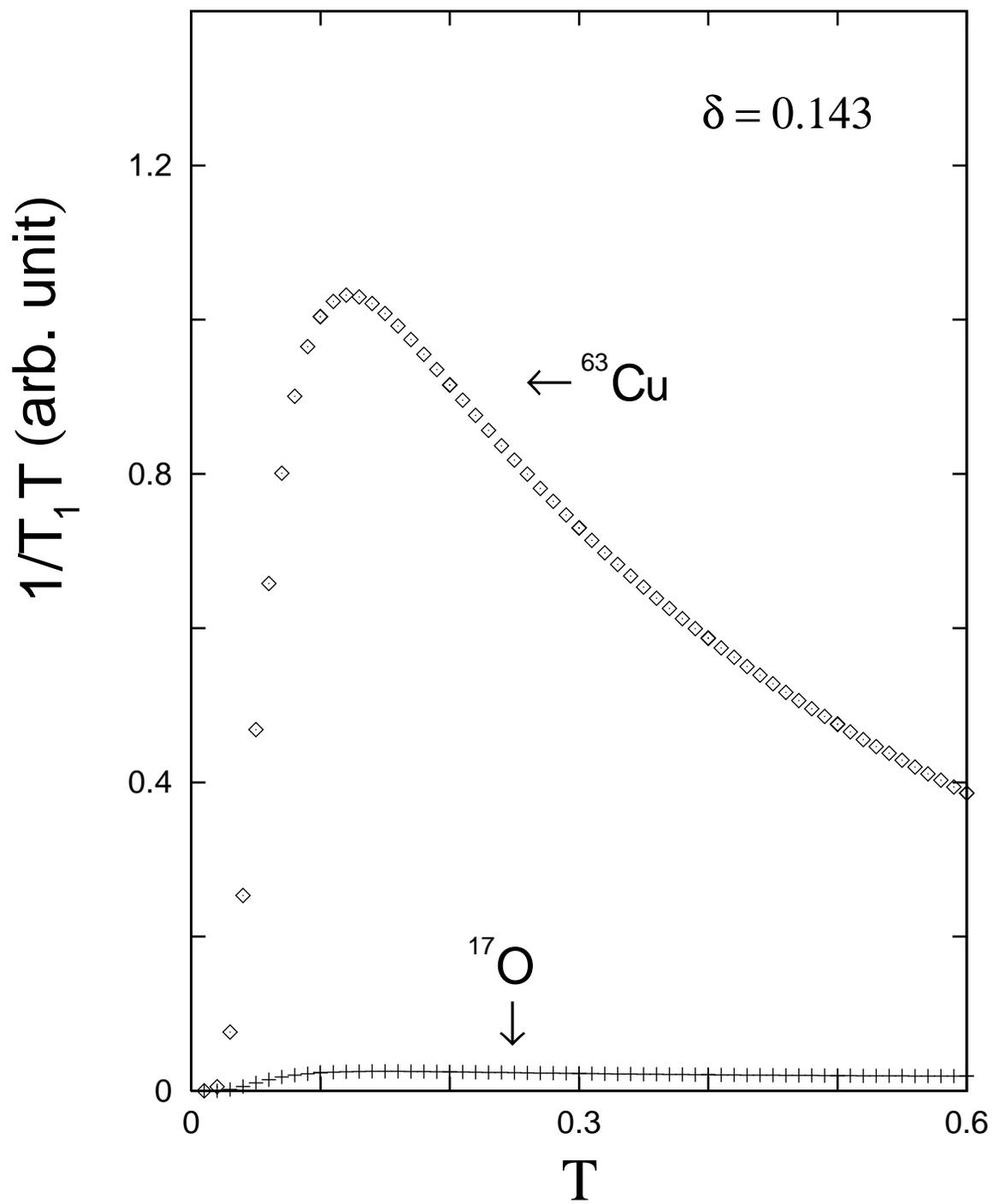
\end{document}